\documentclass[dvips]{acta}
\usepackage{supertabular,lscape,epsfig}
\usepackage{amssymb}
\usepackage{amsmath}
\DeclareSymbolFont{ppa}{OT1}{ppl}{m}{it}
\DeclareMathSymbol{\vv}{\mathalpha}{ppa}{'166}

\newfont{\hb}{rphvb at 10pt}
\newfont{\hbo}{rphvbo at 10pt}
\newfont{\bitt}{rptmbi at 12pt}
\newfont{\bits}{rptmbi at 11pt}

\SetPages{1}{22}

\SetVol{62}{2012}

\begin{document}

\newcommand{\TabApp}[2]{\begin{center}\parbox[t]{#1}{\centerline{
  {\bf Appendix}}
  \vskip2mm
  \centerline{\small {\spaceskip 2pt plus 1pt minus 1pt T a b l e}
  \refstepcounter{table}\thetable}
  \vskip2mm
  \centerline{\footnotesize #2}}
  \vskip3mm
\end{center}}

\newcommand{\TabCapp}[2]{\begin{center}\parbox[t]{#1}{\centerline{
  \small {\spaceskip 2pt plus 1pt minus 1pt T a b l e}
  \refstepcounter{table}\thetable}
  \vskip2mm
  \centerline{\footnotesize #2}}
  \vskip3mm
\end{center}}

\newcommand{\TTabCap}[3]{\begin{center}\parbox[t]{#1}{\centerline{
  \small {\spaceskip 2pt plus 1pt minus 1pt T a b l e}
  \refstepcounter{table}\thetable}
  \vskip2mm
  \centerline{\footnotesize #2}
  \centerline{\footnotesize #3}}
  \vskip1mm
\end{center}}

\newcommand{\MakeTableApp}[4]{\begin{table}[p]\TabApp{#2}{#3}
  \begin{center} \TableFont \begin{tabular}{#1} #4 
  \end{tabular}\end{center}\end{table}}

\newcommand{\MakeTableSepp}[4]{\begin{table}[p]\TabCapp{#2}{#3}
  \begin{center} \TableFont \begin{tabular}{#1} #4 
  \end{tabular}\end{center}\end{table}}

\newcommand{\MakeTableee}[4]{\begin{table}[htb]\TabCapp{#2}{#3}
  \begin{center} \TableFont \begin{tabular}{#1} #4
  \end{tabular}\end{center}\end{table}}

\newcommand{\MakeTablee}[5]{\begin{table}[htb]\TTabCap{#2}{#3}{#4}
  \begin{center} \TableFont \begin{tabular}{#1} #5 
  \end{tabular}\end{center}\end{table}}

\newfont{\bb}{ptmbi8t at 12pt}
\newfont{\bbb}{cmbxti10}
\newfont{\bbbb}{cmbxti10 at 9pt}
\newcommand{\uprule}{\rule{0pt}{2.5ex}}
\newcommand{\douprule}{\rule[-2ex]{0pt}{4.5ex}}
\newcommand{\dorule}{\rule[-2ex]{0pt}{2ex}}
\begin{Titlepage}
\Title{The Optical Gravitational Lensing Experiment.\\
The Catalog of Stellar Proper Motions toward the Magellanic 
Clouds\footnote{Based on observations obtained with the 1.3~m Warsaw telescope at
the Las Campanas Observatory of the Carnegie Institution for Science.}}
\Author{R.~~P~o~l~e~s~k~i$^1$,~~ I.~~S~o~s~z~y~ñ~s~k~i$^1$,~~ A.~~U~d~a~l~s~k~i$^1$,\\ 
M.\,K.~~S~z~y~m~a~ñ~s~k~i$^1$,~~ M.~~K~u~b~i~a~k$^1$,~~ G.~~P~i~e~t~r~z~y~ñ~s~k~i$^{1,2}$,\\ 
£.~~W~y~r~z~y~k~o~w~s~k~i$^{1,3}$~~ and~~ K.~~U~l~a~c~z~y~k$^1$}
{$^1$Warsaw University Observatory, Al.~Ujazdowskie~4, 00-478~Warszawa, Poland\\
e-mail: (rpoleski,soszynsk,udalski,msz,mk,pietrzyn,wyrzykow,kulaczyk)@astrouw.edu.pl\\
$^2$Universidad de Concepción, Departamento de Astronomia, Casilla 160-C, Concepción, Chile\\
$^3$Institute of Astronomy, University of Cambridge, Madingley Road, Cambridge CB3 0HA, UK}
\Received{February 8, 2012}
\end{Titlepage}

\Abstract{
We present a catalog of over 6.2 million stars with measured proper
motions. All these stars are observed in the direction of the Magellanic
Clouds within the brightness range $12<I<19$~mag. Based on these proper
motions about 440\,000 Galactic foreground stars can be selected. Because
the proper motions are based on a few hundred epochs collected during eight
years, their statistical uncertainties are below 0.5~mas/yr for stars
brighter than $I=18.5$~mag. The parallaxes are derived with uncertainties
down to 1.6~mas. For above 13\,000 objects parallaxes are derived with
significance above $3\sigma$, which allows selecting about 270 white dwarfs
(WDs). The search for common proper motion binaries among stars presented
was performed resulting in over 500 candidate systems. The most interesting
ones are candidate halo main sequence star--WD and WD--WD systems. The
application of the catalog to the empirically bound Cepheid instability
strip is also discussed.}{Astrometry -- Catalogs -- Stars: kinematics and
dynamics -- binaries: visual -- globular clusters: individual: 47~Tuc}

\Section{Introduction} 
Proper motions are key parameters for finding intrinsically faint objects
and tracing the dynamical properties of the Galaxy. Both of these
applications require catalogs containing a large number and uniformly
measured proper motions. Most of the existing catalogs of stellar proper
motions are based on a small number of epochs separated by at least 10
years (\eg Fedorov \etal 2009, Girardi \etal 2011). Here we present a
catalog based on a few hundred epochs collected during the third phase of
the Optical Gravitational Lensing Experiment (OGLE-III) project. The epochs
span over 8-yr time baseline. We analyzed the OGLE-III observations in the
directions of the Large and Small Magellanic Clouds (hereafter LMC and SMC;
MCs if both are relevant). The OGLE-III survey observed these dense stellar
regions with the primary goal to search for possible microlensing
events. High stellar density required in order to make the survey
efficient, allowed accurate proper motion and parallax determinations. Our
catalog can be used in those research areas, where most of the other
catalogs of proper motions are less efficient. For the first time the
stellar proper motions in the directions of the MCs are explored in such a
way.

The stars with proper motions $\mu\geq100$~mas/yr were presented separately
(Pole\-ski \etal 2011, hereafter Paper~I). During time baseline of the
OGLE-III these stars changed their position by the amount comparable to the
seeing of the images. Thus, cross-matching their positions from reference
image and distant epoch was harder task and a case-by-case refinement as
good as possible was needed to perform on relatively small number of
objects (\ie 551). These objects are also more important from the
astrophysical point of view.

Here we focus on common proper motion (CPM) binaries. They are composed of
coeval and equidistant stars of the same metallicity, which evolved as
single stars.  CPM systems can be used to study both the stellar evolution
(Monteiro \etal 2006, Sesar \etal 2008, Catal\'an \etal 2008) and the
gravitational potential of the Galaxy (Yoo \etal 2004, Quinn \etal 2009,
Makarov 2012). The other subject which is described in more detail is a
search for the non-variable LMC stars located inside the Cepheid
instability strip. Our high accuracy geometrically measured parallaxes are
used to construct the Hertzsprung-Russell diagram, which allows \eg
selection of white dwarfs (WDs).

Section~2 gives a description of the observations and Section~3 presents
how the catalog was prepared. Section~4 describes the data access. We
evaluate the properties of the catalog in Section~5 and show in Section~6
how it can be used. A summary is presented in Section~7.

\Section{Observations}
\hglue-3pt
The OGLE-III project observed the MCs between 2001 and 2009 with 1.3-m
Warsaw telescope, which is situated at the Las Campanas Observatory,
Chile. The observatory is operated by the Carnegie Institution for
Science. All the observations were collected with the ``second generation''
camera containing eight $2048\times4096$ CCD chips with 15~$\mu$m
pixels. The pixel scale was 0\zdot\arcs26 and total field of view was
$35\arcm\times35\zdot\arcm5$. Only {\it I} and {\it V} filters were used
and about 90\% of observations were secured in the {\it I}-band. The number
of {\it I}-band epochs varied between 385 and 637 for the LMC fields and
between 583 and 762 for the SMC fields with an exception of the field
SMC128 for which 1228 epochs were secured. For the present study we used
{\it I}-band images. {\it V}-band data were used only to provide color
information.  The total sky area observed was 39.8 square degrees for the
LMC fields and 14.2 square degrees for the SMC fields. The sky area
observed was within the Galactic coordinates range
$277\zdot\arcd83<l<283\zdot\arcd87$ and $-38\zdot\arcd07<b<-28\zdot\arcd49$
for the LMC fields. For the SMC fields the ranges were
$299\zdot\arcd35<l<306\zdot\arcd36$ and
$-46\zdot\arcd28<b<-42\zdot\arcd14$. Because of the angular separation of
two groups of fields and different Galactic latitudes, in many cases we
treat them separately. The details of the instrumentation setup were given
by Udalski (2003) while more in-depth description of the observations was
given in Paper~I.

\Section{Catalog Construction}
Most of the published OGLE-III time-series photometry (Udalski \etal 2008a)
was obtained with the Difference Image Analysis (DIA) method. In the
present study we used the time-series astrometry calculated using the {\sc
DoPHOT} software (Schechter \etal 1993). The details of the image alignment
and the {\sc DoPHOT} usage were presented by Udalski \etal (2008a) and in
Paper~I. In Paper~I we also detailed the procedure of proper motion
estimation, which is briefly described below.

Each image taken was spline-resampled to the grid of the reference image of
the given field. For the reference images the transformations from pixel
coordinates to the equatorial coordinates were derived by comparing
positions of bright stars with their 2MASS catalog entries (Skrutskie \etal
2006). The proper motions and parallaxes were derived by the least-squares
fitting of the observed centroid positions ($\alpha$, $\delta$) to the
model defined by the following parameters: proper motion (in right
ascension: $\mu_\alpha$ and in declination: $\mu_\delta$), parallax
($\pi$), differential refraction coefficient ($r$) and equatorial
coordinates for time $t=0$ ($\alpha_0$ and $\delta_0$, $t=0$ corresponds
to the epoch 2000.0):
$$\alpha(t)=\alpha_0+\mu_\alpha t+\frac{r\sin p\tan z+\pi\sin\gamma\sin\beta}{\cos\delta}\eqno(1)$$
$$\delta(t)=\delta_0+\mu_\delta t+r\cos p\tan z+\pi\sin\gamma\cos\beta\eqno(2)$$
where $z$ is the zenith distance, $\gamma$ is the angular distance to the
Sun, $\beta$ is the angle between the direction of the parallax shift and
the direction to the North celestial pole, and $p$ is the angle between the
direction of the refractional shift and the direction to the North
celestial pole.

For each exposure the uncertainty of grid fitting was found based on the
residuals between the measured positions of bright stars and the positions
calculated using the best-fitting models.  This uncertainty was square
added to the uncertainty of the point spread function fitting (Kuijken and
Rich 2002) giving total uncertainty of each position measurement.  The
reciprocal of the square total uncertainty was used as a weight in the
least-squares fitting.  The procedure of fitting parameters was iterative
and in each step the observed positions were corrected so that for stars
brighter than $I=18$~mag the average residua of positions was 0 for a given
exposure and the sigma-clipped average proper motion was equal to 0.  The
dispersion of proper motions for bright stars was used as an estimate of
systematic uncertainties (Paper~I) and for 95\% of subframes was smaller
than 0.67~mas/yr per coordinate.  The zero point of our proper motion
estimates were tied to the LMC, the SMC or 47~Tuc globular cluster (only
for field SMC140).  The absolute proper motions of the MCs and 47~Tuc were
recently discussed by van der Marel \etal (2002), Anderson and King (2003),
Kallivayalil \etal (2006ab) and Piatek \etal (2008).  We present our
measurement of the relative proper motions of 47~Tuc against the SMC in
Section~6.3.

Following the Paper~I, the value of 1.5~mas was square added to the
uncertainties of the parallax measurements. It was found that such a
modification of the formal uncertainties was necessary to properly
reproduce uncertainties derived on the basis of comparison of stars
measured twice -- in the overlapping parts of the adjacent fields.

The proper motions and parallaxes were calculated using this procedure for
all stars in the OGLE-III MCs fields. We were not able to reliably
calculate the proper motions of nearly saturated stars and stars located in
the sky very close to them. For the OGLE-III MCs fields the saturation
limit is between 11.5 and 12.7~mag in the {\it I}-band and depends mostly
on the number density of stars. We removed all objects in the close
proximity of stars brighter than $I=11.5$~mag. We empirically found a
linear relation between {\it I}-band magnitude of bright stars and radius
within which stars had to be removed. It was 1\arcm\ for $I=5.9$~mag down
to 0\arcm\ for $I=11.5$~mag. For a few brightest stars this radius was
found individually. The list of bright stars was taken from the shallow
survey of the LMC performed as a part of the OGLE-III project (Ulaczyk
\etal in preparation) which covered 90\% of the LMC fields. For the rest
of the LMC and all SMC fields we used stars from the DENIS catalog (Cioni
\etal 2000). The areas masked decreased the analyzed sky-area by 0.18 and
0.052 square degrees for the LMC and the SMC, respectively (0.005 square
degrees in the field SMC140).

Apart from the removal of stars neighboring the very bright ones, special
attention was paid to the photometrically variable sources. The changing
flux of a variable star produces a shift in the centroid position, if the
star is blended. If a detached eclipsing binary is blended, the centroids
clump in two groups: one out of eclipse which is closer to the real
position of the star, and second, to which points in the eclipse belong,
much more affected by blending. For some binaries which are heavily blended
and have a small number of points in ingress and egress of the eclipse, the
two groups are fully resolved. In such cases additional change of the
centroid caused by the proper motion or parallax effect are hard to be
distinguished from other factors. That is why we examined these stars
separately in greater details (see Section~6.5). Stars within 1\arcs\ radius
from each variable were removed, which resulted in decreasing the sky area
surveyed by 0.036 and 0.006 square degrees for the LMC and the SMC,
respectively. The list of variable stars was taken from the on-going
research on the OGLE-III Catalog of Variable Stars (hereinafter OIII-CVS)
which includes classical, type II and anomalous Cepheids, RR~Lyr variables,
long period variables, $\delta$~Sct stars, R~CrB variables, eclipsing
binary systems and the so-called double periodic variables (last four types
of variable stars are from the LMC fields only). OIII-CVS was presented in
a series of papers (\eg Soszyñski \etal 2009ab, Graczyk \etal 2011).

Additional candidate variable stars can be selected using the dispersion of
photometric measurements. We did not do this because in the DIA photometry
signi\-ficant changes of the measured flux can be caused by the motion of the
star (Eyer and Wo¼niak 2001). When the distance between current position of
the star and its position on the reference image is comparable to the
diameter of the circle in which the stellar flux is estimated, the measured
flux is smaller than the actual value. Because of the changing distance
between the star and its position on the reference image, the magnitudes
measured using DIA appear to be linear function of time or quadratic
function of time. In this last case the vertex of the parabola corresponds
to the epoch when the actual and reference positions of the star coincided.
For such light-curves the dispersion of photometric measurements can be
significant which hampers clear separation of variable stars from the ones
with high proper motions.
\begin{figure}[p]
\centerline{\includegraphics[width=10.5cm, bb= 30 0 526 410]{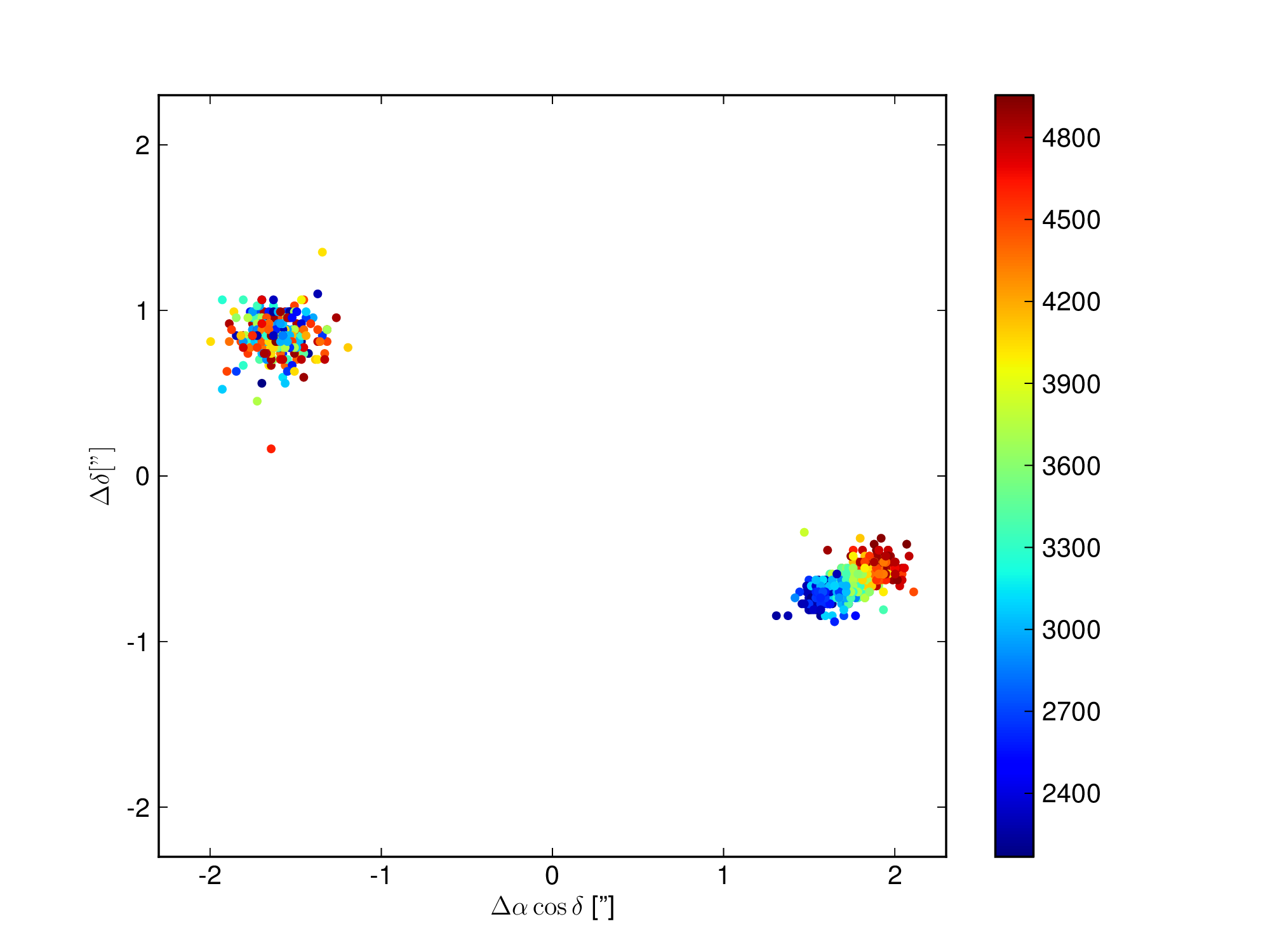}}
\FigCap{Example plot of positions in R.A. and Dec. direction with 
color-coded epoch of observation. The bar shows the color scale of the
epoch of observation ${\rm HJD}-2450000$. The star on the right-hand side
has significant proper motion and the one on the left-hand side is not
moving (\ie belongs to MCs).}
\centerline{\includegraphics[angle=270, width=10.5cm, bb = 22 5 616 774]{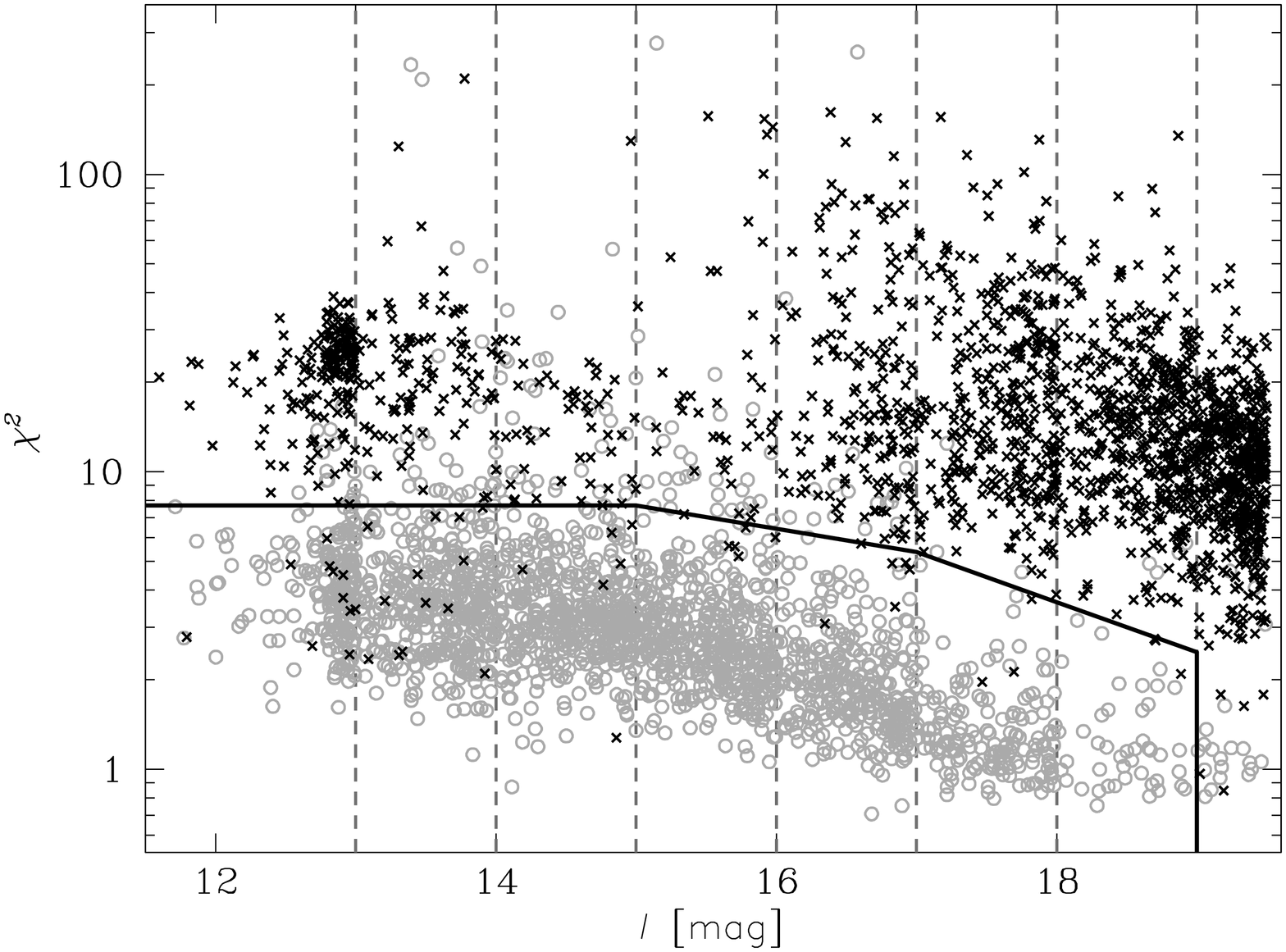}}
\FigCap{$\chi^2$ of Eqs.(1) and (2) model fitting \vs {\it I}-band 
brightness diagram for a subset of stars visually verified. In each 1~mag
range (marked by dashed vertical lines) 500 stars were selected randomly
among objects with $\mu>20$~mas/yr. The gray circles correspond to the
stars which passed the verification and the black crosses -- to the ones
which failed. Black line marks the region which is used for automated
selection of stars with reliable proper motion.}
\end{figure}

The results of fitting the model to the data using the least-squares method
may be significantly changed by outlying measurements, even if their number
is relatively small. In most cases one removes such points using \eg
$3\sigma$ criterion and fits the model once more. It is not obvious how one
can remove outlying points and do not change the derived $\mu$ and $\pi$
for astrophysically important objects (\eg close CPM binaries) at the same
time. Thus, instead of removing outlying points we decided to select stars
with reliably derived proper motions and publish only them.

For selected stars with the proper motions higher than 20~mas/yr we
examined plots of centroids positions in $(\alpha(t),\delta(t))$ plane with
color-coded epoch of the observation $t$. Fig.~1 presents example plot
showing not moving object (left) and a star with significant proper motion
(right). For stars with the proper motions higher than the chosen limit, in
most cases one easily recognizes that either the star is moving or the
proper motion derived is caused by blending with nearby object, small
number of epochs, large scatter of data-points etc. Fig.~2 presents the
diagram of the $\chi^2$ (reduced) of the model fitted \vs {\it I}-band
brightness for a sample of stars. Gray circles mark the stars which passed
the visual examination, while black crosses the ones that failed. Based on
a clear division between the two groups we defined the region on the graph
which gave high reliability of the proper motions. The division is shown by
black line in Fig.~2 and we call it the $\chi^2_{\rm lim}(I)$ criterion.
Even though this region was derived using stars with $\mu>20$~mas/yr, it
was used to automatically select stars with well fitted parameters among
the whole sample of stars.  Table~1 presents, in seven brightness ranges,
reliability of this criterion \ie the number of stars with confirmed proper
motions divided by the number of all stars within the selected area. In
almost all cases it turned out to be higher than 95\%. Table~1 also
presents the completeness which is understood as the ratio of the number of
stars with confirmed proper motions which fulfill the $\chi^2_{\rm lim}(I)$
criterion to the number of all stars for which proper motions could be
measured reliably using our data set.
\MakeTableee{lrr}{6cm}{Statistical properties of criterion applied}
{\hline
\multicolumn{1}{c}{magnitude range}\uprule &
\multicolumn{1}{c}{reliability} &
\multicolumn{1}{c}{completeness} 
\\
\hline
\uprule
$<13$~mag   & 95.1\% & 93.9\% \\
$13-14$~mag & 96.9\% & 92.6\% \\
$14-15$~mag & 97.7\% & 93.2\% \\
$15-16$~mag & 98.9\% & 93.6\% \\
$16-17$~mag & 98.1\% & 95.1\% \\
$17-18$~mag & 98.2\% & 98.2\% \\
$18-19$~mag & 93.9\% & 92.0\% \\
\hline
}

For the stars fainter than $I=19$~mag the scatter of the individual
centroids is so large that examination of our plots did not give firm
conclusions on the proper motion reliability. Thus, we removed all stars
fainter than 19~mag. The stars which did not pass the $\chi^2_{\rm lim}(I)$
criterion were checked, if the fit performed using only centroids from the
images with seeing better than arbitrarily chosen limit of $1\zdot\arcs17$
gave $\chi^2$ below the limit. This way the number of stars fulfilling our
criteria raised by 2.3\%.

The visual examination was performed not only to derive the region on
$\chi^2$ \vs {\it I} plane which corresponded to the well fitted stars. We
also examined the stars of particular interest \ie the ones with the proper
motions higher than 70~mas/yr (no matter whether fulfilling the
$\chi^2_{\rm lim}(I)$ criterion or not), the parallax higher than 15~mas,
candidate WDs and candidate members of CPM binaries with $\mu>30$~mas/yr.
We note that in this step the reliability of the proper motion was checked,
but not the parallax. All the stars that passed the visual examination were
included in the catalog, even if they did not fulfill the $\chi^2_{\rm
lim}(I)$ criterion. The stars which fulfilled this criterion and did not
pass the visual examination were not included to the final catalog.

The parallaxes were estimated for all the stars. If the number of
observations for a given star was small, fitting the model with a smaller
number of parameters should be preferred. Thus, the parallaxes are given
only if the number of measurements was higher than 200 for objects in the
LMC fields and 300 for objects in the SMC fields. The limit was set higher
for the SMC fields, because the distribution of epochs in these fields
hampers parallax estimation. The obvious requirement was that the normal
matrix in the fitting procedure was not ill conditioned. To allow
statistical modeling of the Galaxy kinematics, contrary to Paper I, we did
not put any lower limit on the significance of given parallaxes including
even negative parallaxes.

For simplicity, the coordinates given in the catalog are the same as ones
given by Udalski \etal (2008bc or Paper~I).  We note here that the epochs
to which reference images correspond are between 2001.9 and 2006.9 and can
be found in the headers of the FITS files published by Udalski \etal
(2008bc). The differences between the J2000.0 epoch position and the ones
given by Udalski \etal (2008bc) are in the range 0\zdot\arcs6--0\zdot\arcs7
for only 25 objects presented. The differences are between 0\zdot\arcs5
and 0\zdot\arcs6 for less than 220 stars and for the rest of the stars the
differences are smaller. The OGLE-III fields were overlapping and in some
cases two, three or even four records are given in the catalog for one
star.  We encourage the catalog users to apply their own selection
criteria.

\Section{Data Access}
The catalog of proper motions for stars observed by the OGLE-III survey in
the direction of the MCs is available for the astronomical community only
in the electronic form {\it via} FTP site:
\begin{center}
{\it ftp://ftp.astrouw.edu.pl/ogle/ogle3/pm/mcs/}
\end{center}

The catalog is divided into 1256 files, one for each of the OGLE-III
subfields. For each star we provide: the OGLE-III identifier, J2000.0
equinox coordinates as well as {\it I}-band brightness and $(V-I)$ color
(all taken from Udalski \etal 2008bc or Paper~I), proper motion per
coordinate in both directions ($\mu_{\alpha\star}=\mu_{\alpha}\cos\delta$
and $\mu_{\delta}$) with statistical and systematic uncertainties given
separately (six fields altogether), total proper motion
$\mu=\sqrt{\mu_{\alpha\star}^2+\mu_{\delta}^2}$, parallax and its
uncertainty if available, differential refraction coefficient and its
uncertainty, $\chi^2$ per degree of freedom for the model used,
$\chi^2_{\rm lim}(I)$ value, number of data points used for fitting and
flags showing objects which were visually verified and ones for which only
data from the best seeing images were used. The units of proper motions are
milliarcsecond per year (mas/yr). Both the parallaxes and the differential
refraction coefficients are given in mas. Stars presented in Paper~I are
added for consistency.

Together with the main catalog the files {\sl CPM.dat} and {\sl
variables.dat} are distributed. The first of them describes the CPM systems
found (Section~6.2) and the second one gives the proper motions of the
variable stars (Section~6.5). Remarks to specific objects are given in {\sl
remarks.txt} file. The detailed description of all the catalog files is
given in the {\sl README} file.

\Section{Catalog Properties}
We note that small number of objects in the catalog (1.6\% of the sample)
does not have {\it V}-band magnitude. For these objects and in the field
SMC140 covering central part of 47~Tuc our results are less certain.
 
\subsection{Completeness}
The completeness of our catalog depends on the proper motion value. For
stars published in Paper~I, \ie the ones with $\mu\geq 100$~mas/yr, a very
detailed selection procedure was applied, including case-by-case study of
the time series astrometry. In Paper~I we compared our list with other
published catalogs (Alcock \etal 2001, Soszyñski \etal 2002) showing its
very high completeness. The completeness is only slightly worse for stars
with $70~{\rm mas/yr}\leq\mu<100~{\rm mas/yr}$, as all the stars with the
proper motions in this range were visually verified, even if they did not
pass the $\chi^2_{\rm lim}(I)$ criterion. All the stars with $30~{\rm
mas/yr}\leq\mu<70~{\rm mas/yr}$ passing the $\chi^2_{\rm lim}(I)$ criterion
were also visually verified.  To find this criterion, we inspected also
randomly chosen stars with $20~{\rm mas/yr}\leq\mu<70~{\rm mas/yr}$, which
made their list more reliable and complete than the list of stars with
$\mu<20~{\rm mas/yr}$.

\begin{figure}[htb]
\vglue-6pt
\centerline{\includegraphics[angle=270, width=10.7cm, bb = 22 16 626 832, clip]{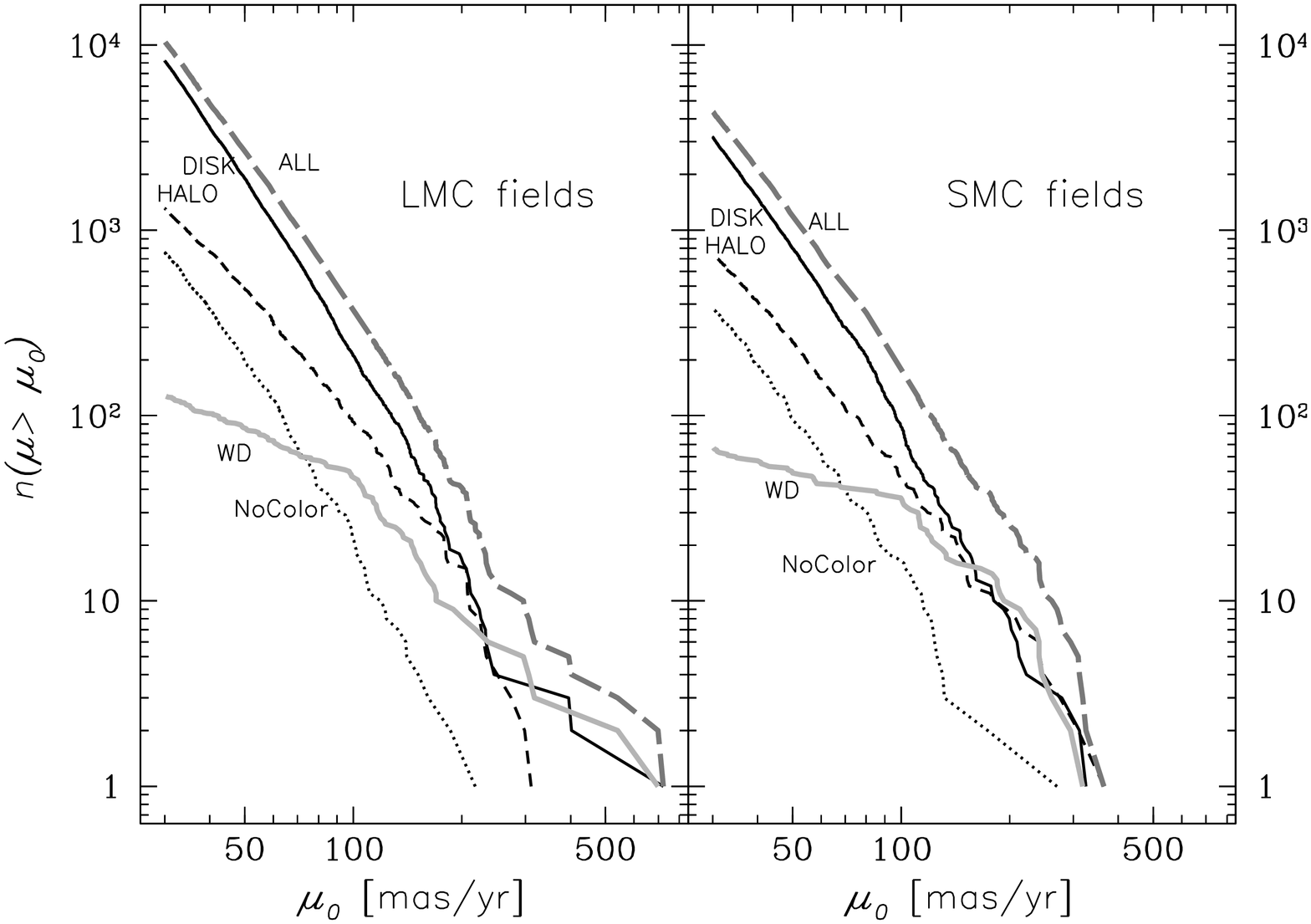}}
\FigCap{Number of objects with proper motions higher than the given limit 
$\mu_0$ for all objects (dashed dark gray line) and separately for: disk
stars (black solid line), halo stars (dashed line), WDs (light gray line)
and the stars without color information (dotted line). {\it Left panel}
presents results for the LMC fields, while right one -- for the SMC
fields.}
\end{figure}

It was checked how the number of stars with $\mu>30$~mas/yr assigned as
disk dwarfs, halo dwarfs, WDs and stars without color information (see
Section~6.1) depends on $\mu$. It is shown in Fig.~3. Only for WD group the
dependence breaks at $\mu\approx100$~mas/yr. We did not find good
explanation for it. In the range $35~{\rm mas/yr}<\mu_0<100~{\rm mas/yr}$
we fitted following relations to the cumulative distribution of all stars
(the LMC and the SMC fields separately):
\setcounter{equation}{2}
\begin{eqnarray}
\log_{10}n_{\rm LMCf}\left(\mu>\mu_0\right) & =  & -2.80\log_{10}\mu_0 + 8.16\\ 
\log_{10}n_{\rm SMCf}\left(\mu>\mu_0\right) & =  & -2.68\log_{10}\mu_0 + 7.62 
\end{eqnarray}
These cumulative distributions were used (see Section~6.2) to statistically
quantify whether the pairs of stars located close to each other on the sky
are highly probable CPM systems or only chance alignments.

The total number of foreground Galactic stars can be estimated by counting
stars with significance of proper motions higher than $5\sigma$. Out of
our 6.2 million stars 440\,000 objects satisfy this criterion.

\subsection{Accuracy of the Proper Motions and Parallaxes}
\begin{figure}[h]
\centerline{\includegraphics[angle=270, width=10.7cm]{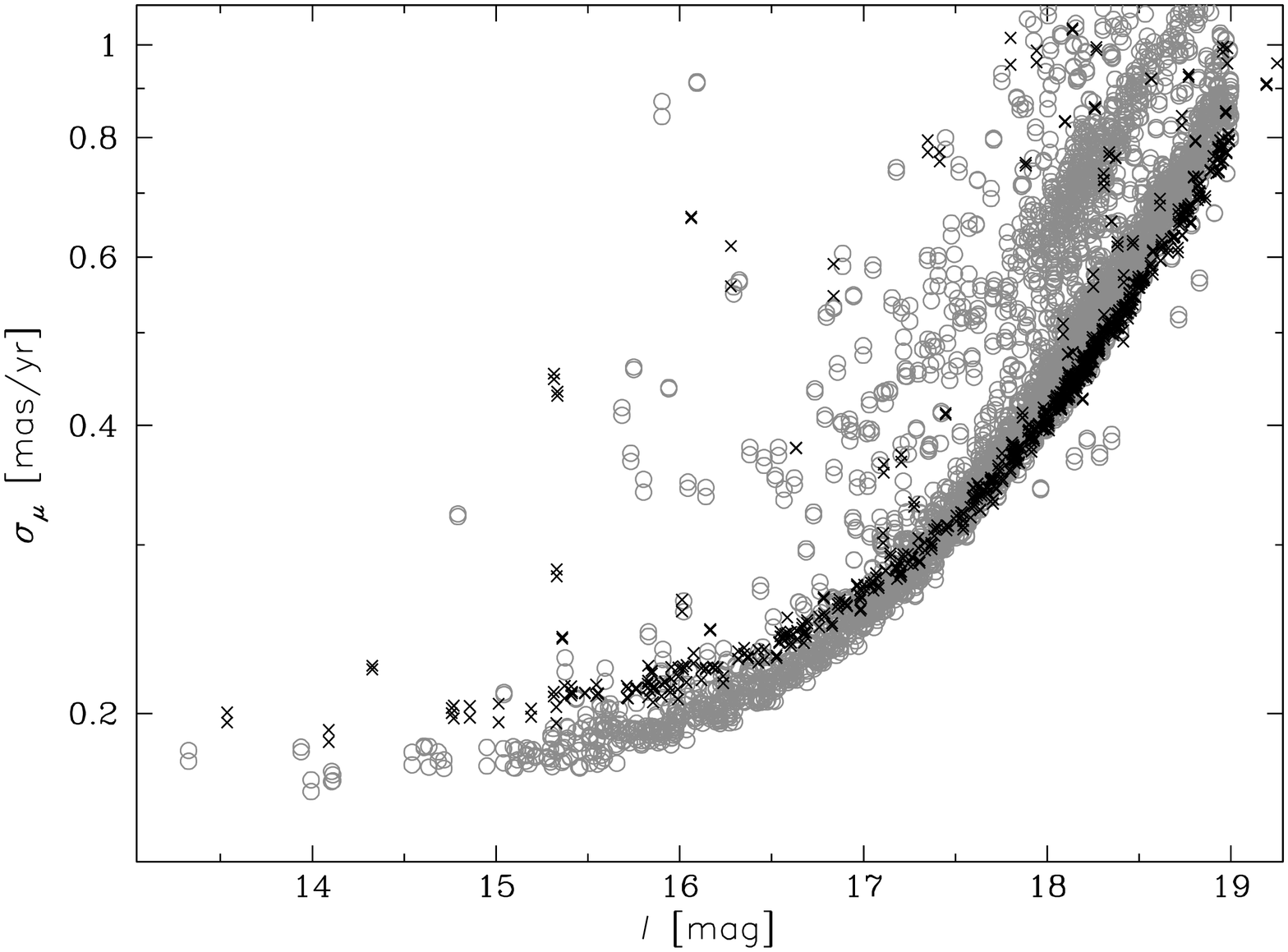}}
\FigCap{Proper motion uncertainties as a function of brightness. 
The crosses correspond to the stars located in a sparse subfield LMC152.1,
while gray circles -- in a dense subfield LMC161.1. Outlying points with
$\sigma_\mu>1.1$~mas/yr are not shown. They compromise around 3\% of the
stars in each field. In the LMC161.1 subfield a group of stars fainter than
18~mag with the uncertainties larger by $\approx30\%$ than most of the
stars with similar brightness is composed of the stars which passed the
selection criterion after removing centroids from images with seeing worse
than $1\zdot\arcs17$. Two points are given for each star depicting
uncertainties in R.A. and Dec. directions separately.}
\end{figure}

Fig.~4 presents the formal uncertainties of the proper motions for stars in
a dense and a sparse stellar fields with similar number of epochs. In the
dense field the uncertainties for the brightest stars are smaller than in
the sparse one because of the higher number of stars used in the frames'
alignment. For stars brighter than 18.5~mag the uncertainties are smaller
than 0.5~mas/yr.

As it was presented in Paper I the uncertainties of the parallaxes are down
to 1.6~mas. The possible effects which may cause systematic offsets in the
parallaxes are too simple description of the differential refraction and
non-zero parallaxes of stars used to calculate corrections of positions.
To calculate these corrections we used stars with $\mu<20$~mas/yr (Paper~I)
but even these stars may influence the zero point of parallaxes comparably
to the smallest uncertainties.

\Section{Discussion}
\subsection{Physical Properties}
\begin{figure}[htb]
\centerline{\includegraphics[angle=270, width=11cm]{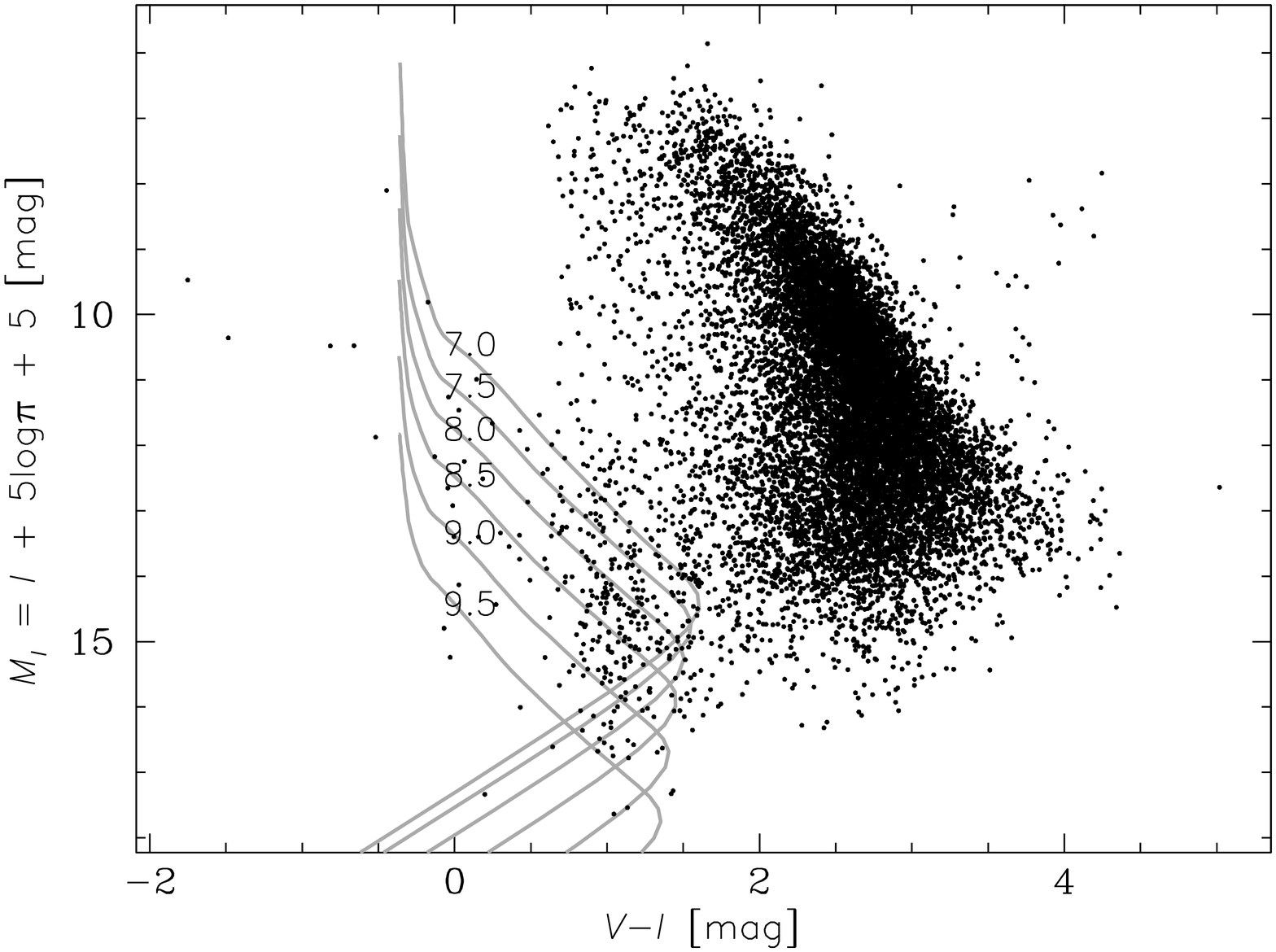}}
\FigCap{Hertzsprung-Russell diagram based on geometric parallaxes. 
Only stars with $\pi/\sigma_\pi>3$ are shown. Error bars are not shown for
clarity. Gray lines present theoretical models for pure Hydrogen WDs by
Holberg and Bergeron (2006) with $\log g$ values given next to each line.}
\end{figure}

The Hertzsprung-Russell diagram for stars with $\pi/\sigma_\pi>3$ is
presented in Fig.~5. The main sequence (MS) is clearly visible.
Theoretical WD positions are plotted based on Holberg and Bergeron (2006)
models. In total, there are $\approx270$ WDs in the part of the diagram
enveloped by the theoretical models. Subdwarfs, which are 2--4~mag fainter
than the MS can also be selected.

The other way, one can verify luminosity class and additionally find
population to which the stars belong, is to plot the reduced proper motion
(hereafter RPM, defined as $H_I=I+5\log\mu$) as a function of the $(V-I)$
color. If all the stars had the same tangential velocities, it would be the
same as H--R diagram shifted vertically by the value depending on the
velocity. Fig.~6 shows the RPM for objects with $\mu>30$~mas/yr. Lines
separating WDs from halo dwarfs (dashed line) and halo dwarfs from disk
dwarfs (dotted line) were determined to give the best discrimination of
different stellar populations. Based on these lines each object was
labeled as WD, halo dwarf, disk dwarf or star without $(V-I)$ color
information. The distribution of stars in our RPM diagram is similar to
the ones derived by other authors (\eg Chanam\'e and Gould 2004, Sesar
\etal 2008).

\begin{figure}[htb]
\vglue-12pt
\centerline{\includegraphics[angle=270, width=11cm]{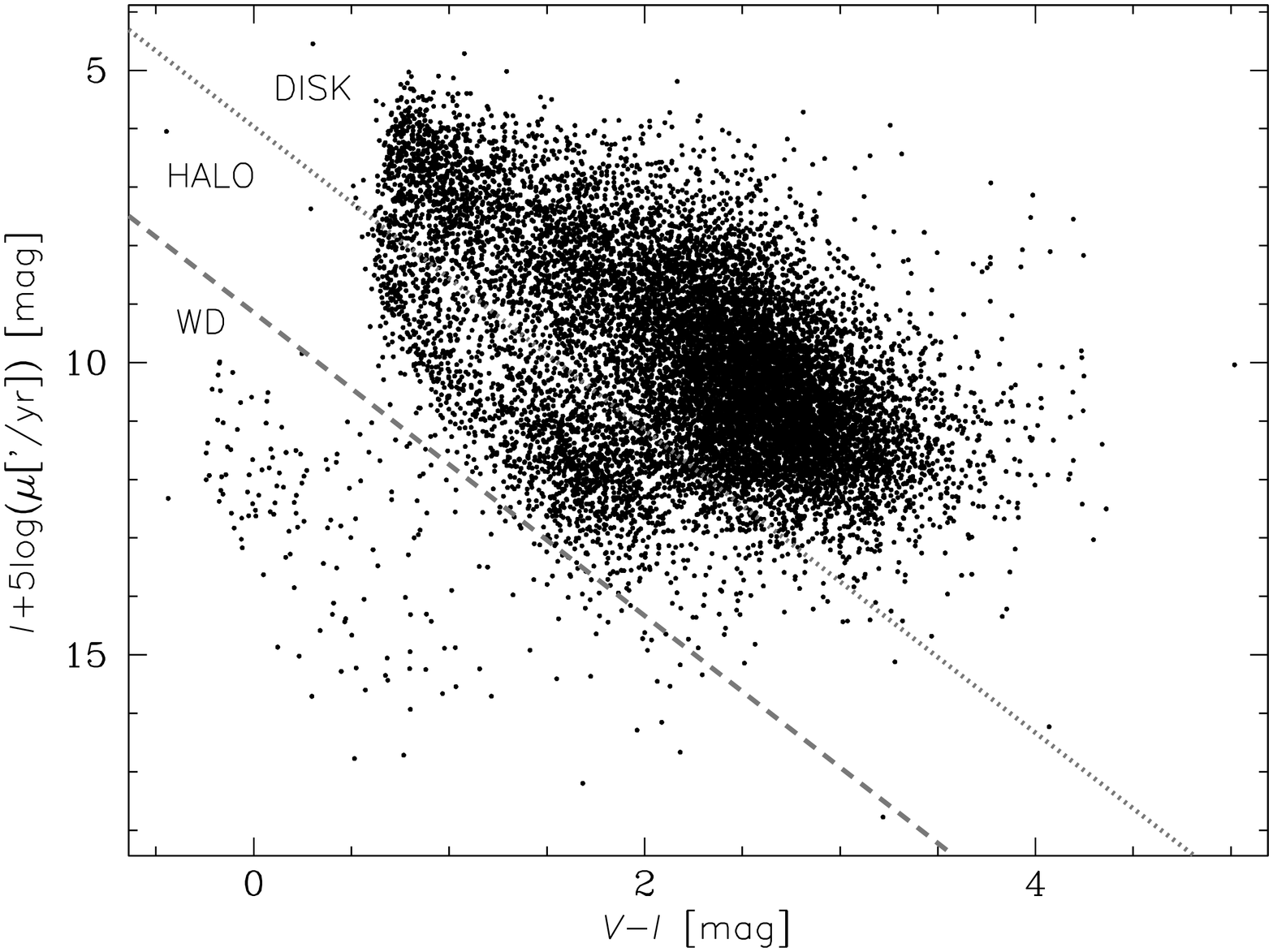}}
\FigCap{Reduced proper motion diagram for all stars with $\mu>30$~mas/yr.
Dashed line separates WDs from halo stars while dotted line separates halo
stars from disk stars.}
\end{figure}

\subsection{Common Proper Motion Systems}
To search for the CPM binaries we used the stars with $\mu>\mu_{\lim}=
30$~mas/yr. The edges of the OGLE-III fields were overlapping, thus, before
the search was started a list of unique stars with $\mu>\mu_{\lim}$ was
prepared. It contained 10\,405 stars located in the LMC fields and 4378
stars located in the SMC fields.  All of them were inspected visually, not
only to check the reliability of the results, but also to find the CPM
companions at the separations around 1\arcs. In such close systems the
mean magnitudes of components may be inaccurately measured. For some stars
we manually removed outlying points and fitted the model once more.

All the stars were searched for companions closer than 1500\arcs. Each pair
was verified similarly to Chanam\'e and Gould (2004). First, by comparing
the angular distance of the components ($\Delta\theta$) and the vector
proper motion difference ($\Delta\mu$) with a distribution of unrelated
pairs. These distributions were normalized based on the number of stars
with proper motions higher than the proper motion of the pair. We report
all the CPM binaries which passed this selection. Second, it was checked on
the RPM diagram if both components belong to the same population. We
describe these two steps below and present the most interesting systems.

\begin{figure}[htb]
\centerline{\includegraphics[width=7.99cm]{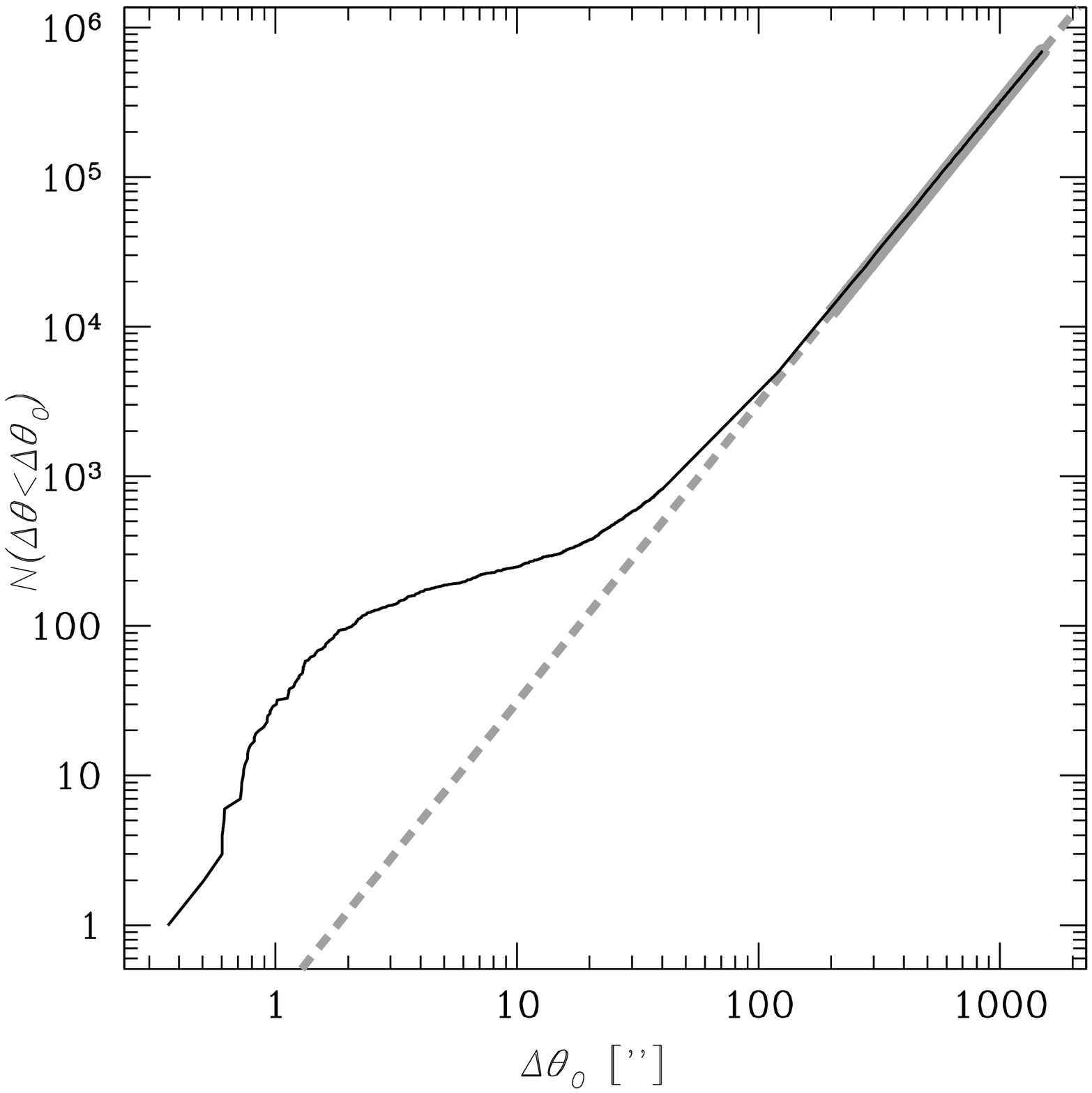}}
\FigCap{Cumulative distribution of number of pairs as a function of angular 
separation for all pairs of stars with $\mu>30$~mas/yr in the LMC fields
with $\Delta\theta<1500\arcs$. The gray line shows the relation for
unrelated pairs. Thick part corresponds to the region where the fit was
performed and dashed part is an extrapolation.}
\end{figure}

The expected number of unrelated pairs ($N_{\rm UP}$) with angular
separations greater than the value for a given pair ($\Delta\theta_0$), the
vector proper motion difference ($\Delta\mu$) smaller than the value for a
given pair ($\Delta\mu_0$), and which are found among stars with proper
motions larger than for a given pair ($\mu_0$), was estimated based on:
$$N_{\rm UP}(\Delta\theta_0,\Delta\mu_0,n(\mu>\mu_0))=N(\Delta\theta<\Delta\theta_0,n(\mu>\mu_0))p(\Delta\mu<\Delta\mu_0)\eqno(5)$$
where $n(\mu>\mu_0)$ is the number of stars with proper motion larger than
$\mu_0$ in the LMC fields (Eq.~3) or in the SMC fields (Eq.~4),
$N(\Delta\theta<\Delta\theta_0,n)$ denotes the number of unrelated pairs
with separations smaller than $\Delta\theta_0$ found among $n$ stars, and
$p(\Delta\mu<\Delta\mu_0)$ is the probability that an unrelated pair has
the proper motion vector difference ($\Delta\mu$) smaller than
$\Delta\mu_0$.

Fig.~7 presents the cumulative number of pairs (black line) as a function
of $\Delta\theta$ based on all pairs with separations $\Delta\theta <
1500\arcs$ found among 10\,405 stars in the LMC fields.  Genuine CPM
binaries are expected to have small values of $\Delta\theta$, and the
excess of binaries with $\Delta\theta \lesssim 20\arcs$ is evident. It is
expected that the number of unrelated pairs increases as a square of
$\Delta\theta_0$.  Very small number of pairs with $\Delta\theta$ in the
range between 200\arcs\ and 1500\arcs\ are expected to be CPM binaries.
The total of over 680\,000 pairs found in this range results in the fit
$$N(\Delta\theta<\Delta\theta_0)=0.31~\Delta\theta^2_0$$
The gray line in Fig.~7 presents this relation. The slope of observational
data is very close to the expected one in the range of $\Delta\theta$ were
the fit was found. From this fit we found that the separation within which
one unrelated pair is expected is as small as 1\zdot\arcs8.

The number of unrelated binaries scales also as a square of the number of
stars used for search ($n$). The above search was performed with
$n=10\,405$. This allows to rewrite the above equation for $N$ taking into
account $n$:
$$N(\Delta\theta<\Delta\theta_0, n)=2.9\cdot10^{-9}~n^2\Delta\theta^2_0\eqno(6)$$

\begin{figure}[htb]
\centerline{\includegraphics[angle=270,width=12.5cm, bb= 202 0 596 809, clip]{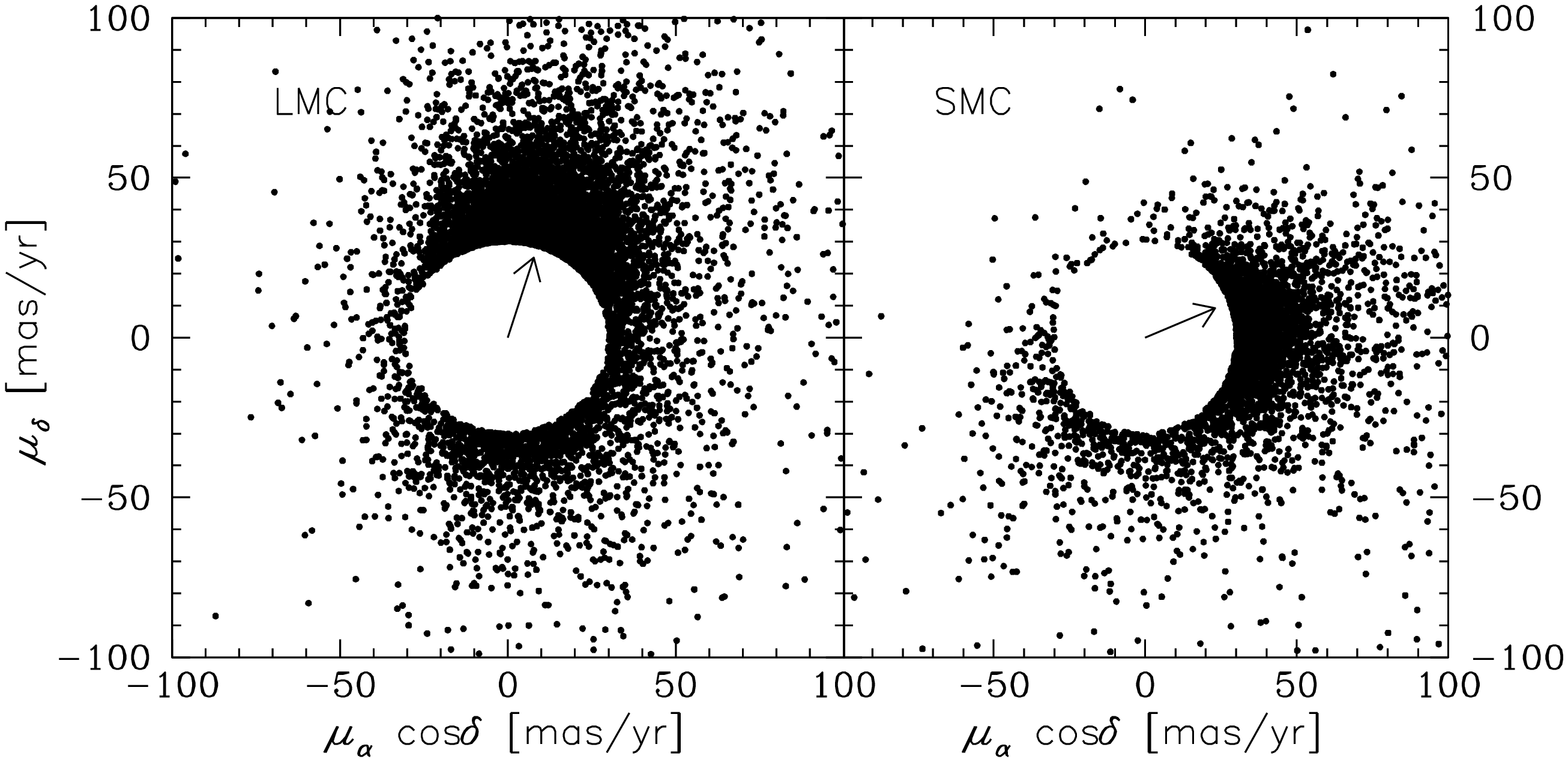}}
\FigCap{Proper motion vector-point diagram for stars in the LMC fields
({\it left panel}) and the SMC fields ({\it right panel}). Only stars with
$\mu>\mu_{\lim}$ are shown.  The figures are limited to the absolute values
of proper motions per coordinate below 100~mas/yr.  The arrows indicate the
average direction of the solar anti-apex based on Sch\"onrich \etal (2010).
The position angle of solar anti-apex direction for individual star may
deviate from the average up to 18\zdot\arcd5 in the LMC fields and up to
9\zdot\arcd5 in the SMC fields.}
\end{figure}

\begin{figure}[htb]
\centerline{\includegraphics[width=9cm]{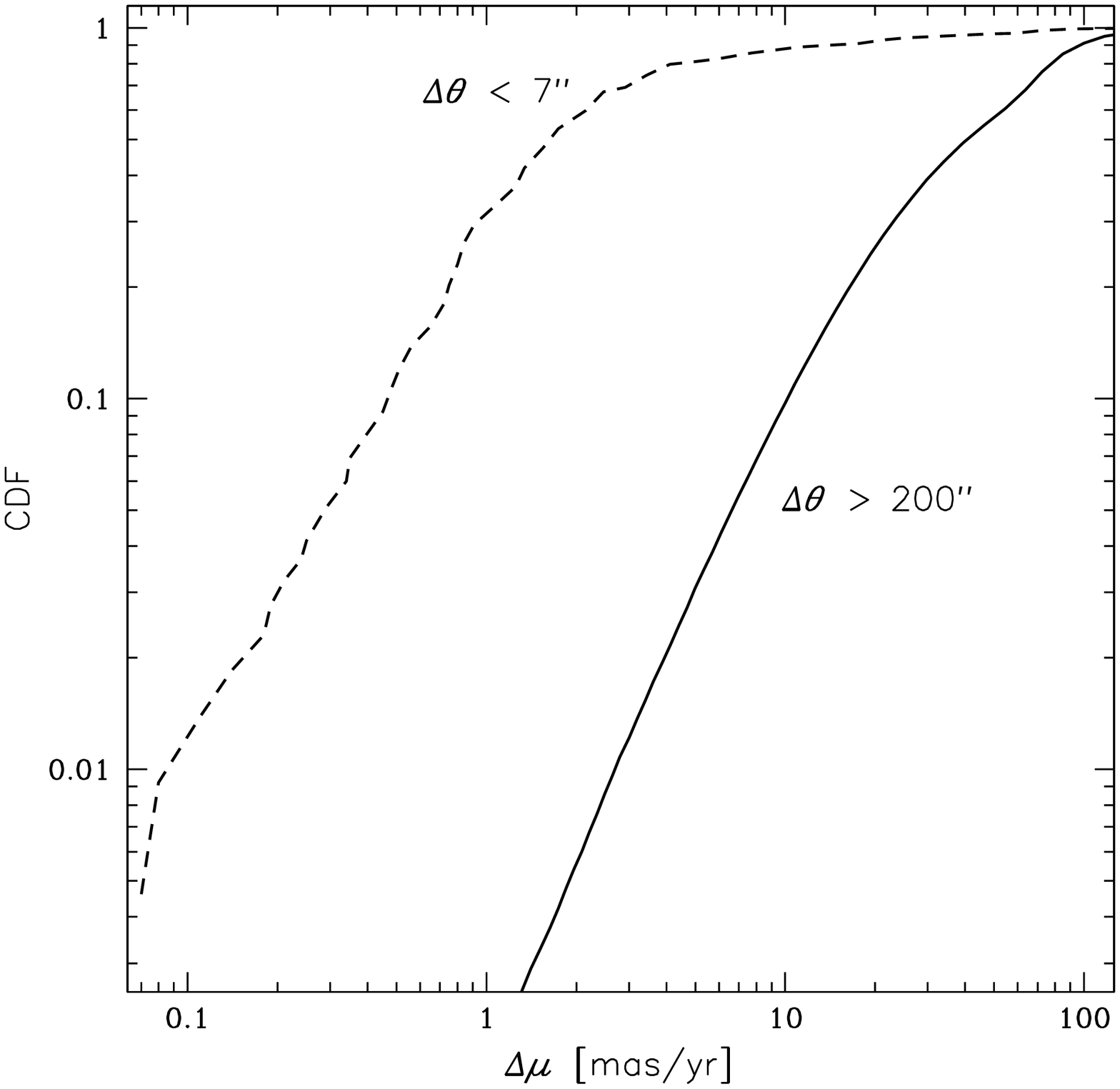}}
\FigCap{Cumulative Distribution Functions (CDF) of $\Delta\mu$ for the
pairs in the LMC fields with separations $\Delta\theta>200\arcs$ (\ie
unrelated pairs; solid line) and $\Delta\theta<7\arcs$ (\ie mostly CPM
binaries; dashed line).}
\end{figure}

To estimate $p(\Delta\mu<\Delta\mu_0)$ probability one can assume that
$\mu_{\alpha\star}$ and $\mu_\delta$ are random variables taken from a
distribution with known properties. In such a case the formula for
$p(\Delta\mu<\Delta\mu_0)$ could be found based on theoretical basis with
the parameter fitting as we did for $N(\Delta\theta<\Delta\theta_0,
n)$. Fig.~8 presents the proper motion vector-point diagram for stars with
$\mu>30$~mas/yr and located in the LMC fields (left panel) and the SMC
fields (right panel). Even without the $\mu_{\lim}$ limit these
distributions could not be well approximated by bivariate Gaussian
distributions. That is because the vectors of solar anti-apex are not the
axes of reflection symmetry of these distributions. Proper description of
such distributions using distributions with known properties would be very
challenging task. Thus, to estimate $p(\Delta\mu<\Delta\mu_0)$ we used the
cumulative distribution functions (CDFs) of $\Delta\mu$ for pairs with
$\Delta\theta>200\arcs$. As it was mentioned before, almost all of these
pairs are expected to be unrelated. Solid line in Fig.~9 shows the CDF of
$\Delta\mu$ for the pairs in the LMC fields with $\Delta\theta>200\arcs$.
For comparison we plot on the same figure CDF of $\Delta\mu$ for pairs with
$\Delta\theta<7\arcs$ (dashed line) which are mostly expected to be CPM
binaries. The much faster increase of the dashed line compared to the solid
line, proofs that closer pairs have components of similar proper
motion. The CDF of pairs with $\Delta\theta>200\arcs$ in the SMC fields
overlays the one for the LMC fields and is not plotted for clarity.

For each pair fulfilling the criteria $\Delta\theta<1500\arcs$ and
$\Delta\mu<30$~mas/yr, we calculated $N_{\rm UP}(\Delta\theta_0,
\Delta\mu_0,n(\mu>\mu_0))$. The $n(\mu>\mu_0)$ parameter was
estimated based on Eqs.~(3) or (4) for the LMC and the SMC fields,
respectively. Eq.~(6) was used to calculate $N(\Delta\theta<
\Delta\theta,n)$ and $p(\Delta\mu<\Delta\mu_0)$ probability was
empirically found based on the CDF of $\Delta\mu$ (Fig.~9). Most of the CPM
binaries should have small values of $N_{\rm UP}$.  We decided to search
for pairs with $N_{\rm UP}(\Delta\theta_0,\Delta\mu_0,n(\mu>\mu_0))<0.75$,
which resulted in 316 pairs in the LMC fields and 214 pairs in the SMC
fields. All of them are reported in the file {\sl CPM.dat}. The sum of
$N_{\rm UP}(\Delta\theta<\Delta\theta_0,
\Delta\mu<\Delta\mu_0,n(\mu>\mu_0))$ values for all reported pairs
resulted in $53.8$, which is the expected number of unrelated pairs in the
selected sample.

Next, we verified whether for pairs composed of halo or disk stars their
components have positions on the RPM diagram consistent with the positions
expected for coeval objects belonging to the same population. As noted by
Chanam\'e and Gould (2004) the components of CPM binary should lie on the
line parallel to the MS track. Each pair was flagged as either consistent,
inconsistent or questionable. There are 70 pairs containing WDs or stars
without color information, which are not characterized in this way. Also
for 98 pairs containing stars located very close to each other on the sky
or on the RPM diagram verification is not possible. For veri\-fied objects
there were 256 (71\%) marked as consistent, 71 (20\%) as inconsistent and
35 (9.7\%) as questionable.

\begin{figure}[htb]
\centerline{\includegraphics[angle=270, width=9cm, bb = 0 16 596 784, clip]{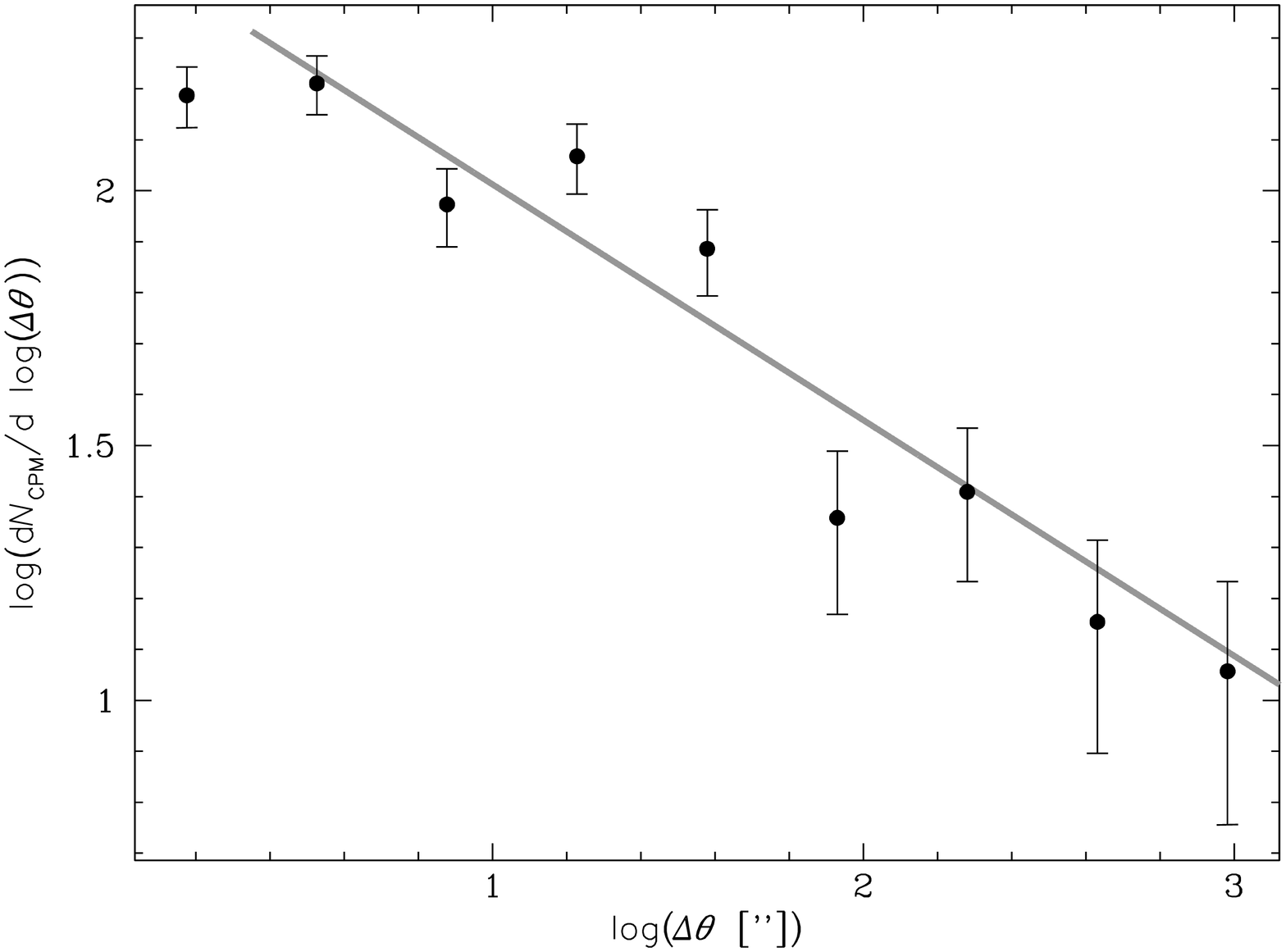}}
\FigCap{Distribution of disk CPM binaries angular separations. 
The error bars were calculated assuming Poisson statistics.}
\end{figure}

Fig.~10 presents the distribution of the disk CPM binaries (confirmed using
the RPM diagram) as a function of angular separation. All the points except
the one for the smallest $\Delta\theta$ were used to make a linear fit
(gray line). If we assume that the number of CPM ($N_{\rm CPM}$) scales as
$N_{\rm CPM}\approx\theta^{-\alpha}$ then the inclination of the
best-fitting line gives the estimated power-law exponent $\alpha$ of
$1.46\pm0.02$.  This value is significantly smaller than the $1.67\pm 0.07$
value found by Chanam\'e and Gould (2004). We suspect this discrepancy is
caused by the stellar streams which we observe in the narrow solid angle
analyzed here. The impact of stellar streams is reduced in the whole sky
analysis such as the one by Chanam\'e and Gould (2004). The median parallax
of the CPM binaries used in a fit is 4.4~mas which translates to the
typical distance almost four times larger than in Chanam\'e and Gould
(2004) sample. Only the first point in Fig.~10 deviates significantly from
the linear fit, thus, the completeness of our sample of CPM binaries seems
constant in the range from 2\arcs\ to $\approx 1000\arcs$. The number of
confirmed halo CPM binaries is too small to perform an analysis of the
angular separations' histogram.

One candidate pair containing a WD and a halo star \ie LMC155.1.4867 and
LMC155.1.5999 was announced in Paper~I. Monteiro \etal (2006) observed a
similar pair to estimate the age of the red subdwarf. Similar research can
be done using LMC155.1.4867-LMC155.1.5999 pair, in which additionally the
distance of both components is larger, brightness difference is smaller,
and brighter component does not affect observed spectrum of the fainter
component, as was a case for Monteiro \etal (2006) observations. Other
interesting pair from Paper~I contains two WDs (LMC102.7.22769 and
LMC102.7.22886). The separation of these objects is 92\arcs. There are
only 32 other WD-WD CPM systems known (Sion \etal 1991, Jordan \etal 1998,
Scholz \etal 2002, Farihi \etal 2005 and reference therein). The
LMC107.2.14205-LMC107.3.195 pair with $\Delta\theta=593\arcs$ was
confirmed on the RPM diagram. If parallaxes of the components ($9.2\pm
1.6$~mas and $9.4\pm1.6$~mas, respectively) are taken into account, this
angular separation translates into the projected physical separation of
0.31~pc, which is the largest value among the RPM-confirmed systems
presented here and with significant parallaxes of both components.

It is possible to search and verify in detail CPM binaries among stars with
lower proper motion limit. The number of candidate CPM binaries, selected
based on the expected number of unrelated pairs, scales as
$\mu^{-2.5}_{\lim}$. For a search performed on the list of stars with
$\mu_{\lim}=20$~mas/yr, we can expect a total of around 1460 candidate CPM
binaries. The preparation of the reliable list of such binaries requires
much of additional effort and is beyond the scope of this paper.

\subsection{Relative Proper Motion of 47 Tuc and SMC}
The relative proper motion of the Galactic globular cluster 47~Tuc and SMC
was measured in the two subframes of the SMC136 field and averaged. These
subframes contained a reasonably high number of bright stars ($I<17.7$~mag)
belonging to both environments -- 112 in the SMC and 113 in 47~Tuc.
Table~2 lists the literature values of the relative proper motion and
includes our estimate. The measurements of absolute proper motion of 47~Tuc
were transformed to relative values based on the absolute proper motion of
the SMC from Piatek \etal (2008) \ie $\mu_{{\rm SMC},\alpha\star}=0.754
\pm0.061$~mas/yr and $\mu_{{\rm SMC},\delta}=-1.252\pm0.058$~mas/yr.
The only measurement which is significantly more accurate than ours is
based on the Hubble Space Telescope (HST) observations (Anderson and King
2003).

\MakeTableee{lrr}{6cm}{Relative proper motions of 47~Tuc and SMC}
{\hline
\multicolumn{1}{c}{Source}\uprule &
\multicolumn{1}{c}{$\mu_{\alpha\star}$ [mas/yr]} &
\multicolumn{1}{c}{$\mu_{\delta}$ [mas/yr]} 
\\
\hline
\uprule
Tucholke (1992)			& $5.5 \pm 2.0$		& $-1.6 \pm 2.0$ \\
Odenkirchen \etal (1997)$^a$	& $6.2 \pm 1.0$		& $-4.05 \pm 1.0$ \\
Freire \etal (2001)$^a$		& $5.8 \pm 1.9$ 	& $-2.15 \pm 0.6$ \\
Freire \etal (2003)$^a$		& $4.5 \pm 0.6$		& $-2.05 \pm 0.6$ \\
Anderson and King (2003) 	& $4.716 \pm 0.035$ 	& $-1.357 \pm 0.021$ \\
Girard \etal (2011)$^{ab}$	& $6.9 \pm 1.0$		& $-2.1 \pm 1.0$ \\
this work 			& $4.41 \pm 0.67$ 	& $-1.12 \pm 0.55$ \\
\hline
\noalign{\vskip4pt}
\multicolumn{3}{p{9cm}}{$^a$ -- shifted based on the absolute proper motion
of the SMC by Piatek \etal (2008); $^b$ -- uncertainty anticipated by the authors is given;} 
}

\subsection{Absolute Proper Motion of the Magellanic Clouds}
There are 200 quasars known behind the LMC (Koz³owski \etal 2012). Most
of the brightest ones are close to stars of similar brightness which
hampers the determination of the proper motion with the highest possible
accuracy. Our data do not permit the determination of the proper motion of
the LMC or the SMC against quasars with accuracy comparable to the HST one
(Piatek \etal 2008).

\subsection{Proper Motions of Variable Stars}
Proper motions of variable stars for which absolute magnitudes can be
estimated using other methods (\eg RR~Lyr variables) can be used to analyze
statistical properties of their kinematics and distribution (\eg Martin and
Morrison 1998, Kinman \etal 2007). To allow similar analysis, we separately
present proper motions for stars from OIII-CVS.

As discussed earlier, measuring the proper motion of a variable star is
more complicated than for a constant star. All the variable stars for which
at least one model resulted in a significant proper motion were visually
inspected similarly to the method shown in Fig.~1. We paid special
attention to check, if for these stars the observed motion was not caused
by the light variations on a very long time-scales. In such cases the
apparent motion might be a result of blending only. No significant
parallaxes for the variable stars were found.

There are three possible astrophysical situations in which the proper
motion of a variable star observed toward the MCs may be significant: (i)~a
variable star is located in the foreground of the MCs, (ii)~a variable star
in the MCs is blended with a foreground object, which causes the centroid
of both objects to move and (iii)~variable star in the MCs has an
exceptionally high tangential velocity and it is possibly a runaway object.
For some of the low signal-to-noise variables (\eg $\delta$~Sct pulsators
and smallest amplitude long period variables) it is possible that the
claimed variability is caused by the photometric noise. The situation
(i) can in most cases be verified based on the position of the object in
the color--magnitude diagram (CMD) or the period--luminosity diagram. The
situation (ii) can be verified using high spatial resolution imaging. The
last possibility is quite unlikely and if confirmed, would reveal very rare
objects.

A group of variable stars to which we pay special attention were blue
variables from LMC. Sabogal \etal (2005) claimed that in the LMC these
stars separate into two groups in the CMD. One group contains stars redder
than $(B-V)=0.4$~mag. Most of those stars have $-0.3~{\rm mag}<(B-V)<
0.0$~mag. The second group, which has not been seen in the SMC, contains
stars with the colors in the range of $0.4~{\rm mag}<(B-V)<0.6$~mag. We
have cross-matched these stars with our catalog of proper motions and it
turned out that the stars from the second group are not blue variables from
the LMC but Galactic stars which are observed in the direction of the
LMC. In the second group 87.5\% of stars had proper motions larger than
4~mas/yr, while in the first one this number was only 0.5\%. Most probably
the brightness changes seen in the photometry were caused only by the
change of the stellar centroids (Alcock \etal 2001, Eyer and Wo¼niak
2001). Our finding is confirmed by the color-color diagram shown by Sabogal
\etal (2005) in which the stars from the second group follow main sequence
track of spectral types later than F5.

The variable stars with significant proper motions are presented in a
separate file {\sl variables.dat}. It contains OIII-CVS identifier,
equatorial coordinates and proper motions ($\mu_{\alpha\star}$,
$\mu_\delta$ and $\mu$). In total 236 variable stars are presented. Most of
them (162) are eclipsing binaries. Three CPM binaries containing variable
stars are indicated in {\sl remarks.txt} file distributed with the catalog.

\subsection{Cepheid Instability Strip}
The catalog of proper motions can be used to check if a given set of stars
belong to the MCs or the Galaxy. Below, we present one of possible
applications -- a search for non-variable stars in the LMC which are
located in the part of the CMD populated by the classical Cepheids
(Soszyñski \etal 2008). Foreground stars are major contaminants in this
part of the diagram and we removed them using proper motions, which is
described below.
\begin{figure}[p]
\centerline{\includegraphics[angle=270,width=12.5cm, bb = 202 30 626 754, clip]{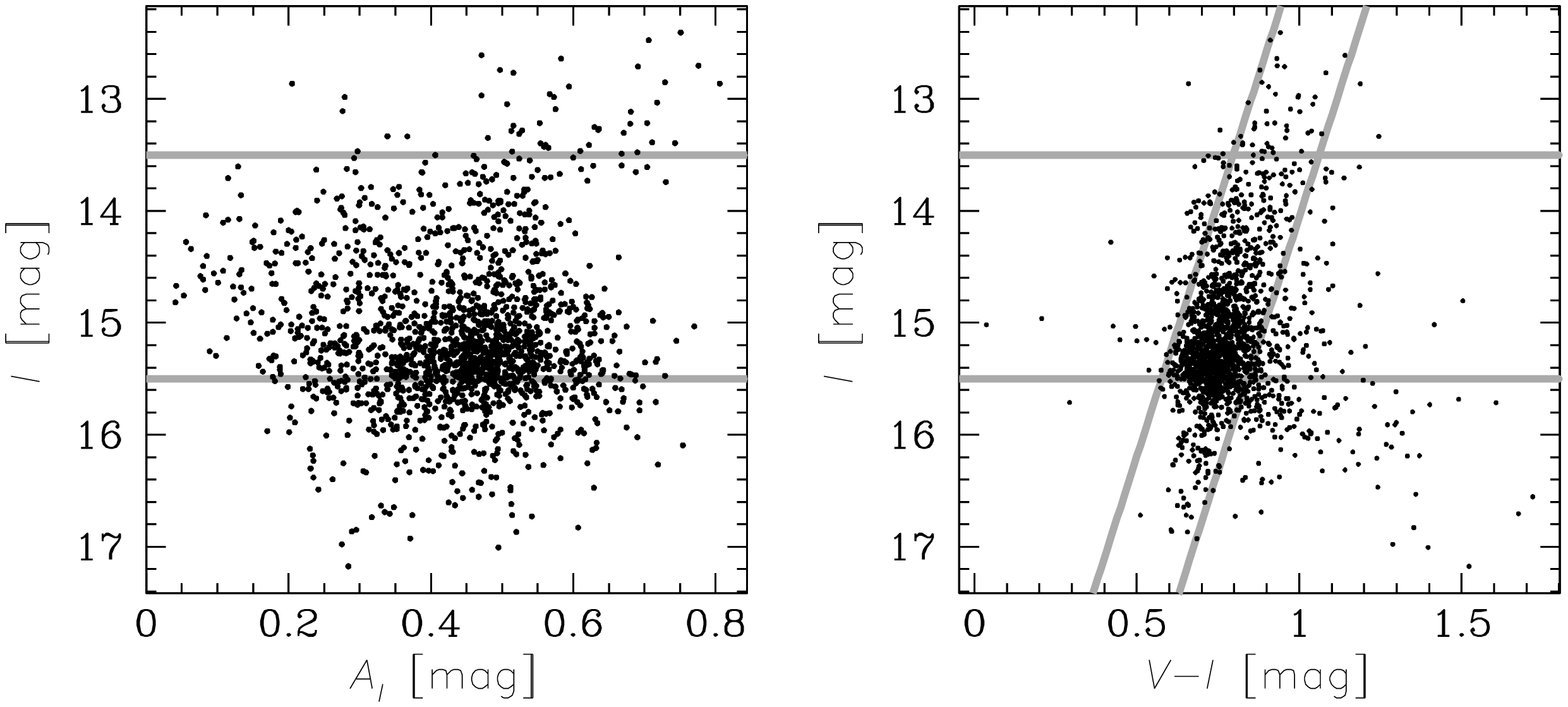}}
\FigCap{{\it Left panel}: {\it I}-band brightness \vs {\it I}-band 
amplitude for fundamental mode Cepheids in the LMC. {\it Right panel}: CMD
of those Cepheids. Gray lines constrain the part of the CMD from which
constant stars are further analyzed.}
\end{figure}
\begin{figure}[p]
\centerline{\includegraphics[width=12.5cm]{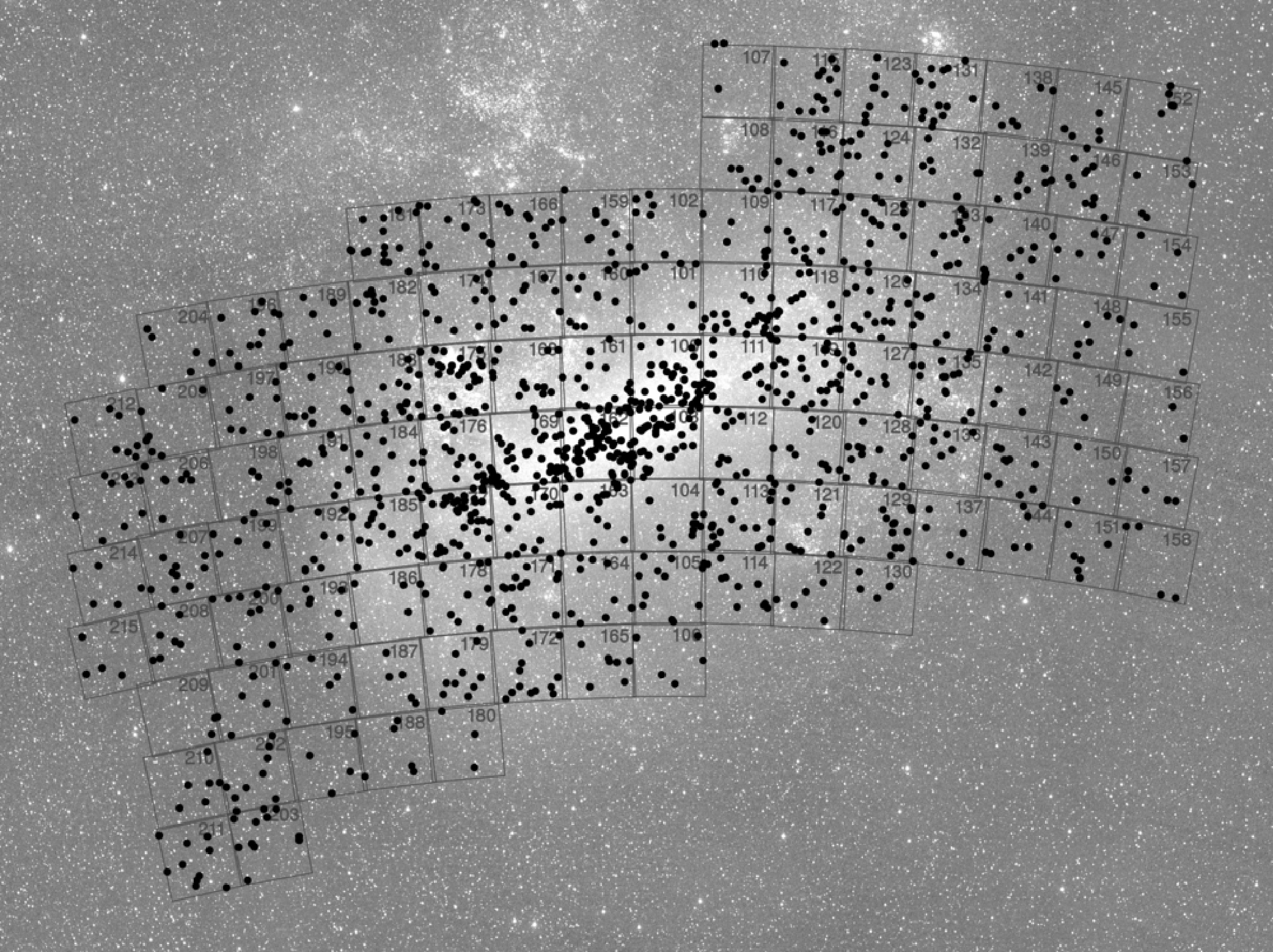}}
\FigCap{Sky projection of constant objects with proper motions consistent 
with the LMC and located inside the region defined in the right panel of
Fig.~11. Background image comes from the ASAS survey (Pojmañski 1997).
The squares represent the OGLE-III fields with field number given for each
of them.}
\end{figure}

The left panel of Fig.~11 presents the mean {\it I}-band brightness \vs
{\it I}-band amplitude for fundamental mode Cepheids in the LMC. The right
panel of Fig.~11 shows the CMD of these Cepheids.  As we wanted to restrict
our search to the region of CMD where Cepheids are efficiently found and
high accuracy proper motions are available, we limited brightness range to
$13.5~{\rm mag}<I<15.5~{\rm mag}$ (shown on both panels of Fig.~11) and
used objects lying between two inclined lines on the CMD (shown on the
right panel of Fig.~11). These four lines constrain the part of the CMD
from which stars are further analyzed. Except the Cepheids there are
14\,840 stars in our selection region for which the proper motions were
obtained from the catalog. In order to remove Galactic objects from the
sample we selected stars with proper motions consistent with the LMC ones
($\mu<2$~mas/yr) and consistent with zero within $2\sigma$ limit. In total
1361 stars fulfilled all the above constraints. Their positions are shown
in Fig.~12 with the LMC image from the ASAS survey in the background. The
selected objects clearly clump in the bar of the LMC, which is another
strong suggestion that most of them belong to that galaxy.

The blending scenario can be verified for some of these objects using the
existing HST observations. Ground-based Str\"omgren photometry of these
objects was already obtained in selected fields. When analyzed it will give
the final verification if these objects are dwarfs or white dwarfs and thus
belong to the Galaxy, or are giants and supergiants and thus belong to the
LMC. The latter case will allow observational mapping of the Cepheid
instability strip.

\Section{Summary}
The presented catalog gives proper motions for stars observed in the
directions of the LMC and the SMC. For bright stars the accuracy is below
1~mas/yr including systematic uncertainties. Uncertainties of the derived
parallaxes are down to 1.6~mas.

Beside presenting the catalog itself, we outlined a few ways it can be
used. Geometrically measured parallaxes were used to construct the
Hertzsprung-Russell diagram, which permits the selection of $\approx270$
WDs. The RPM diagram enabled the discrimination of the stars belonging to
the Milky Way halo from disk objects. Among the stars with the proper
motion $\mu>30$~mas/yr we searched for the CPM binary systems. Based on
statistical considerations more than 500 candidate CPM binaries were
selected. Among pairs which could be verified on the RPM diagram 7.7\%
showed positions inconsistent with any isochrone. The distribution of the
RPM-confirmed disk CPM binaries is well fitted with a power law exponent of
$1.46\pm0.02$, which is significantly smaller than the $1.67\pm0.07$ value
found by Chanam\'e and Gould (2004). Among CPM binaries one candidate for a
WD-WD pair and one candidate for a halo star-WD were found. In order to
assess the accuracy of our measurements, we determined the proper motion of
47~Tuc globular cluster and compared it with other estimates. Measurements
significantly better than ours were obtained only using the HST. We payed
special attention to the variable stars for which proper motions were
presented separately.  The existence of blue LMC variables with $(B-V)>
0.4$~mag, claimed by Sabogal \etal (2005), was neglected.

The catalog may be used to prove or disprove that selected stars or groups
of stars belong to the LMC or the SMC. As an example, we selected the
objects which are located inside the classical Cepheid instability strip on
CMD, did not show significant light variations in the OGLE-III data and
have proper motions coherent with the LMC one. Additional observations will
then allow empirical bounding of the instability strip.

\Acknow{Authors thank S. Koz³owski for comments on early draft and 
J.~Chanam\'e for discussion. This work was supported by MNiSW grant
N-N203-512538. RP was supported through Polish Science Foundation START
program. The OGLE project has received funding from the European Research
Council under the European Community's Seventh Framework Programme
(FP7/2007-2013)/ERC grant agreement No.~246678.}

\end{document}